\documentclass{article}
\usepackage{amssymb}
\newcommand{\wick}[1]{\protect{:\!#1\!:}}
\def\1ad{\mbox{\normalsize $^1$}}
\def\2ad{\mbox{\normalsize $^2$}}
\def\3ad{\mbox{\normalsize $^3$}}
\def\4ad{\mbox{\normalsize $^4$}}
\def\5ad{\mbox{\normalsize $^5$}}
\def\6ad{\mbox{\normalsize $^6$}}
\def\7ad{\mbox{\normalsize $^7$}}
\def\8ad{\mbox{\normalsize $^8$}}
\def\makefront{
\vspace*{1cm}\begin{center}
\def\sp{
\renewcommand{\thefootnote}{\fnsymbol{footnote}}
\footnote[4]{corresponding author : \email_speaker}
\renewcommand{\thefootnote}{\arabic{footnote}}
}
\def\newtitleline{\\ \vskip 5pt}
{\Large\bf\titleline}\\
\vskip 1truecm
{\large\bf\authors}\\
\vskip 5truemm
\addresses
\end{center}
\vskip 1truecm
{\bf Abstract:}
\abstracttext
\vskip 1truecm
}
\def\beq{\begin{equation}}                     %
\def\eeq{\end{equation}}                       %
\def\bea{\begin{eqnarray}}                     
\def\eea{\end{eqnarray}}                       
\begin {document}                 
\def\titleline{
Unitary Quantum Field Theory 
\newtitleline
on the Noncommutative Minkowski space   
}
\def\email_speaker{
{\tt 
bahns@mail.desy.de       
}}
\def\authors{
D. Bahns\sp
}
\def\addresses{
II. Institut f\"ur Theoretische Physik\\
Universit\"at Hamburg\\Luruper Chaussee 149\\
D - 22761 Hamburg
}

\def\abstracttext{ This is the written version of a talk I gave at the 35th
Symposium Ahrenshoop in Berlin, Germany, August 2002. It is
an exposition of joint work with S.~Doplicher,
K.~Fredenhagen, and Gh.~Piacitelli~\cite{bdfp}. The violation of unitarity
found in quantum field theory on noncommutative spacetimes in the context of
the so-called modified Feynman rules is linked to the notion of time ordering
implicitely used in the assumption that perturbation theory may be done in
terms of Feynman propagators. Two alternative approaches which do not entail a
violation of unitarity are sketched. An outlook upon our more recent work is
given. }

\large
\makefront


\section{Introduction}

Noncommutative spacetimes are studied for various reasons, one of them being
the {\em Ge\-dan\-ken\-experi\-ment} that Heisenberg's uncertainty relation
along with the laws of classical gravity leads to a restriction as to the best
possible localization of an event in spacetime. The idea is that the
simultaneous measurement of two or more spacetime directions with an
arbitrarily high precision would require an arbitrarily high energy which in
the end would result in forming a horizon, cf. for instance~\cite{dfr}. Another
motiviation is based on string theory~\cite{csw}. 

The mathematical model on which our analysis is founded was defined
in~\cite{dfr}, where continuous spacetime is replaced by a noncommutative
*-algebra $\mathcal E$ generated\footnote{To be exact, I should mention that the
noncommutative coordinates are merely affiliated to the noncommutative algebra 
in the sense of Woronowicz, but in this talk I shall
skip all technical details. See~\cite{dfr}.} 
by Hermitean noncommutative coordinate-operators
$q_0,\ldots,q_3$ with $[q^\mu,q^\nu]=iQ^{\mu\nu}$ subject to  {``quantum
conditions''},
\beq
Q_{\mu\nu} Q^{\mu\nu}\,=\,0
\,,\qquad
\big({\textstyle\frac{1}{4}}\,Q_{\mu\nu}\,Q_{\rho\sigma}\,
\epsilon^{\,\mu\nu\rho\sigma}\big)^2 
\,=\,\lambda_P^8 \, I
\,,\qquad
[q_\rho,Q_{\mu\nu}]\,=\,0
\eeq
where $\lambda_P$ is the Planck length. The quantum conditions are Poincar\'e
invariant, and entail that for any state
$\omega$ in the domain of the $[q_\mu,q_\nu]$, the uncertainties $\Delta_\omega
q_\mu = \sqrt{\omega (q_\mu^2) - \omega (q_\mu)^2}$ fulfill the following 
space-time uncertainty relations~\cite{dfr},
\beq
\Delta q_0 \cdot \left (\Delta q_1 +
\Delta q_2 +\Delta q_3\right)  \,\geq\, \lambda_P^2
\,,\qquad
\Delta q_1\cdot\Delta q_2 + \Delta q_1\cdot\Delta q_3  + \Delta q_2 \cdot\Delta
q_3\,\geq\, \lambda_P^2
\eeq 
Note that while it is obvious that for
spacelike noncommutativity the first of the above uncertainty relations is
trivial, it is not clear whether for lightlike noncommutativity~\cite{mehen},
where $Q_{\mu\nu} Q^{\mu\nu}\,= \,Q_{\mu\nu}\,Q_{\rho\sigma}\,
\epsilon^{\,\mu\nu\rho\sigma} \,=\,0$, such uncertainty relations hold at all,
since in the above framework both right hand sides would be zero,
cf.~\cite[p.199]{dfr}.

The regular realizations of the quantum conditions lead to the Weyl-Wigner
calculus of ordinary quantum mechanics. In particular, the product in~$\mathcal
E$ is given by the twisted convolution, 
\beq\label{twcon}
f(q) g(q) = \int d^4k \,d^4l\,\check f (k) \,\check g(l)\,e^{-\frac{i}{2}kQl}
\,e^{i(k_\mu+l_\mu) q^\mu}
= \int d^4k \,\,e^{i k_\mu q^\mu}
\,(f * g)\check{\phantom{l}}(k)
\eeq
with $f\in \mathcal F L^1(\mathbb R^4)$, $\check f = {\mathcal F}^{-1} f$, 
where $\mathcal F$ is the ordinary Fouriertransform, and
\beq\label{twpos}
f*g(x)=\int dx_1dx_2\,e^{2i\,(x-x_1)\,Q^{-1}\,(x_2-x)}\,f(x_1)\,g(x_2)
\eeq
In the literature (\ref{twcon}) is often thought of as being defined by the
inverse Fourier transform of the Moyal star product instead of the
product~(\ref{twpos}). 

It should also be stressed that in the approach followed here, $Q$ is
not a fixed matrix, but that by the quantum conditions, the joint spectrum
$\Sigma$ of the operators~$Q_{\mu\nu}$ is homeomorphic to the non-compact
manifold $TS^2 \times \{1,-1\}$ and that $\mathcal E$ is a trivial bundle over
$\Sigma$. The full Poincar\'e group acts as automorphisms on $\mathcal E$, and
derivatives may be defined as the infinitesimal generators of translations.
Note also that the evaluation in a point $f(q)\rightarrow f(a)$ is not a
positive functional on $\mathcal E$. For all details see~\cite{dfr}.


\section{Perturbation Theory}

In~\cite{dfr} the free field on $\mathcal E$ was formally defined as
\beq
\phi(q) := \frac{1}{(2\pi)^{3/2}}
\int \frac{d^3\vec k}{2 \,k_0}
\left.\left( a(k)\otimes e^{-ikq} + a^*(k)\otimes e^{ikq}\right)
\right|_{k_0=\sqrt{{\vec k }^2 + m^2}}
\,\,=\int d^4k \,\check\phi(k)\,e^{ikq}
\eeq
where $\phi(q+x):=U(x)\phi(q)U(x)^{-1}$ is to be understood as an operator
valued distribution, $f\mapsto \int dx\, f(x) \phi(q+x)$. From~(\ref{twcon}) we
deduce that products of fields are nonlocal, rendering the perturbative
definition of an interacting quantum field theory difficult, since
it is for instance far from obvious how the time ordering
or time-zero fields should be defined. Moreover, no equivalent of
Osterwalder-Schrader-positivity has yet been proved, and it is unclear how a
Euclidean version of the theory could be related to a quantum field theory in
the Minkowski regime.\footnote{In fact, since I gave this talk, we have found
that the Euclidean approach and the one on Minkowski space cannot be easily
related as certain tadpoles which are finite in the Euclidean regime cease to
be so on Minkowski space.} Let us now consider the so-called modified Feynman
rules~\cite{filk} as one possible approach to treating interactions on
noncommutative spacetimes.

\subsection{Modified Feynman rules}
Starting point of this approach is the action functional
\beq
S[\phi]=  {\int d^4q}
 \,\Big(\mathcal L_0 (q) +\lambda \phi(q)^n\Big)
= \int d^4x \,\Big(\mathcal L_0 (x) +\lambda \underbrace{\phi * \dots
* \phi}_n  (x)  \Big) 
\eeq
where $\int d^4 q$ is the trace on $\mathcal E$. In the literature, the twisted
convolution (or the Moyal star product) is often taken at a particular point
$\sigma\in\Sigma$, and thereby Lorentz invariance is broken explicitely. 

Since $\int d^4q
\,f(q)g(q)=\int d^4x  \,f g(x)$, the free action is the same as in ordinary
quantum field theory, whereas the interaction is nonlocal, and given by
\beq
\lambda\int d^4k_1 \dots d^4k_n \, \hat\phi (k_1) \dots
 \hat \phi(k_n)  \, 
  {e^{-\frac{i}{2}\sum\limits_{i<j}k_i\sigma k_j}}
 \delta^{(4)}({\textstyle\sum}\, k_i) 
\eeq
It is therefore tempting to {\em assume} that the ordinary perturbative setup
can be used, where  Feynman propagators serve as internal lines. The 
nonlocality of the interaction is then taken into account by simply adding
adequate twisting factors $\exp\big({-\frac{i}{2}\sum_{i<j}k_i\sigma k_j}\big)$
at the vertices. 

Unfortunately, as was shown in~\cite{mehen}, this setup leads to a violation of
unitarity, since in $\phi^3$-selfinteracting theory at second order perturbation
theory the optical theorem does not hold, i.e.
$2$ Im 
\begin{picture}(40,10)(0,8)
\put(00,10){\line(3,0){10}}
\put(10,10){\circle{2}}
\put(30,10){\line(3,0){10}}
\put(30,10){\circle{2}}
\put(20,10){\circle{20}}
\end{picture}
$\neq$
$\left|\right.$
\begin{picture}(20,10)(0,8)
\put(00,10){\line(3,0){10}}
\put(10,10){\line(3,2){10}}
\put(10,10){\line(3,-2){10}}
\put(10,10){\circle{2}}
\end{picture}
$\left.\right|^2$,
unless spacelike or lightlike noncommutativity is assumed.

In~\cite{bdfp} we have linked this phenomenon to the definition of
time-ordering by comparing the modified Feynman rules with two alternative
approaches which independently of the chosen quantum
conditions do not entail a violation of unitarity.


\subsection{Hamiltonian approach}

The first of these approaches was already given in~\cite{dfr} and is based on
the introduction of a Hamiltonian,
\beq
H (t) =  {\int\limits_{q_0=t} d^3\vec q}
  \,\, \mathcal H (q)
= \int\limits_{x_0=t} d^3\vec x
\,\Big(\,\mathcal H_0(x) + \mathcal H_I (x)\,\Big)
\eeq
where $\int_{q_0=t} d^3 \vec q$ is defined as a positive weight on $\mathcal E$, 
$\mathcal H_0 (x)$ is the  ordinary free Hamiltonian
and the interaction Hamiltonian is given in terms of the twisted convolution, 
$\mathcal H_I (x) = \lambda \,\wick{\phi* \dots * \phi(x)}$. As in ordinary
field theory, the corresponding $S$-Matrix is then defined as a formal
power series in the coupling constant,
\beq
S = I + \sum\limits_{r=1}^\infty\frac{(-i)^r}{r!}\displaystyle\int dt_1 \dots
dt_r\,\mbox{\bf\large 
T} H(t_1)\dots H(t_r)\, 
\eeq
where the time ordering $\mbox{\bf\large T} H(t_1)\dots H(t_r)$ is defined with
respect to the {\em parameter times} $t_1,\dots,t_r$, and therefore separated
from the nonlocal products. Since the Hamiltonian is
symmetric, i.e. $H(t)^*=H(t)$, the $S$-matrix is obviously unitary (up to
possible violations arising in the renormalization procedure).

The explicit difference between this approach and the modified Feynman rules was
sketched in~\cite{bdfp}. For instance, at second order in  $\phi^3$-interaction,
the Hamiltonian approach would yield the following contribution to the
$S$-matrix,
\bea
&&\int dt_1 dt_2 \,\theta(t_1-t_2) \hspace{-1.7ex}
\int\limits_{x_0=t_1} \hspace{-1.7ex}d^3 \vec x \hspace{-1.3ex} 
\int\limits_{y_0=t_2} \hspace{-1.7ex}d^3 \vec y \,
\,\wick{\phi(x)^{* 3}}\,\,
\wick{\phi(y)^{* 3}}\,\,
\\&+&
\int dt_1 dt_2 \,\theta(t_2-t_1) \hspace{-1.7ex}
\int\limits_{y_0=t_2} \hspace{-1.7ex}d^3 \vec y \hspace{-1.3ex} 
\int\limits_{x_0=t_1} \hspace{-1.7ex}d^3 \vec x \,
\,\wick{\phi(y)^{* 3}}\,\,
\wick{\phi(x)^{* 3}} 
\eea
with the Heaviside function $\theta$.
In order to calculate expectation values of the above explicitely,  we apply
the Wick theorem and pick up all possible contractions, using ordinary formulas
such as $\langle \Omega |\check \phi (k_1) \check\phi(k_2)
|\Omega\rangle=(2\pi)^{-4}\,\hat \Delta_+(k_2) \delta (k_1+k_2)$. One
immediately finds that some contractions will involve only pointwise products,
while some will involve twisted convolutions. The former are referred to
as planar contributions while the latter are called nonplanar.

The important observation then is that contrary to the assumption made
in the context of the modified Feynman rules we do {\em not} have Feynman
propagators in terms which involve twistings. For instance, we find  
the following nonplanar contribution to the fish graph: $\theta\cdot\Delta_+\star 
\Delta_+ +(1-\theta)\cdot\Delta_-\star \Delta_-$ where the symbol $\star$ is
used instead of $*$ to indicate that the twisted convolution is to be taken
with respect to $2Q$. While the corresponding planar graph
$\theta\cdot\Delta_+^2 +(1-\theta)\cdot\Delta_-^2$ indeed yields the square of
the Feynman propagator, $\Delta_F^2$, this is {\em not} the case for the
nonplanar contribution,
\beq
\theta\cdot\Delta_+\star  \Delta_+
+(1-\theta)\cdot\Delta_-\star \Delta_-
\neq  \Delta_F\star \Delta_F
\eeq
unless the Heaviside function $\theta$ may pass the twisted convolution, in the
sense that \linebreak$\theta\cdot\Delta_\pm\star  \Delta_\pm=
\theta\Delta_\pm\star\, \theta \Delta_\pm$. This is not generally true, but
may be done only in the case of spacelike or lightlike
noncommutativity. In these cases, the assumption
that Feynman propagators serve as internal lines is compatible with the requirement that the theory
be unitary~\cite{bdfp}.
For additional explicit calculations in this framework see~\cite{sibold}.


\subsection{Yang Feldman equation}

A covariant approach to perturbation theory on the noncommutative Minkowski
space was given in~\cite{bdfp}. It is based on the field equation and results
in a direct perturbative definition of the interacting field~\cite{YFK}. As
early as 1952 this approach was already used in the analysis of nonlocal field
theories~\cite{KM}. The idea is to solve the field equation
\beq
(\Box+m^2)\phi(q)=-\lambda
\phi(q)^{n-1}
\eeq
perturbatively by
\beq
\phi(q)=\sum_{k=0}^\infty\lambda^k\phi_k(q)
\eeq
Identifying $\phi_{0}(q)$ with the incoming field $\phi_{in}(q)$, we have at
$k$-th order
\beq
\phi_k(q)=\int dy \,\,\,\Delta_{ret} (y)
\hspace{-3ex}\sum_{k_1+\dots k_{n-1}=k-1}\hspace{-3ex}\phi_{k_1} (q-y)\cdots
\phi_{k_{n-1}}(q-y)
\eeq
with {\em ordinary} retarded propagators $\Delta_{ret}$. As in the Hamiltonian
approach, the time ordering is thus separated from the nonlocal products. The
graph theory of the above construction is given by rooted trees with $n-1$
branches at each vertex and with retarded propagators connecting different
vertices. Unitarity in this context means that the interacting field must be
Hermitean, which it obviously is if we assume that the incoming field is
Hermitean.

As an explicit example let us again consider $\phi^3$-theory, where we have
only 2 branches at each vertex, such that at first order, the interacting field
is 
\beq
\phi_1(q)\quad=\quad
\int dy \,\,\, {\Delta_{ret}} (y) \, 
\phi_{0} (q-y) \phi_{0}(q-y)
\,
\begin{picture}(60,10)(0,7)
\put(30,25){\circle{2}}
\put(50,25){\circle{2}}
\put(36,0){ {\line(0,0){15}}}
\put(30,25){\line(1,-1){10}}
\put(40,15){\line(1,1){10}}
\put(45,12){$\leftarrow q-y$}
\put(45,0){$\leftarrow q$}
\end{picture}
\qquad
\eeq
and at second order,
\bea
\phi_2(q)&=&\int dy \,\,\, {\Delta_{ret}}
 (y)\,(\,\phi_{0} (q-y) \phi_{1}(q-y) 
+\phi_{1} (q-y) \phi_{0}(q-y) \,)
\\&=&\int dy \,\,\, {\Delta_{ret}} (y)\,\int dz
\,\,\, {\Delta_{ret}} (z) \,\Big(
\,\phi_{0} (q-y) \,\phi_{0}(q-y-z)\,\phi_{0}(q-y-z)
\\&&
\phantom{\int dy \,\,\, {\Delta_{ret}} (y)\int dz
{\Delta_{ret}} (z)}
+\,\,\phi_{0} (q-y-z)\,\phi_{0} (q-y-z) \,\phi_{0}(q-y) \,\Big)\qquad
\\
&=&
\begin{picture}(60,30)(-15,5)
\put(40,35){\circle{2}}
\put(60,35){\circle{2}}
\put(30,25){\circle{2}}
\put(40,35){\line(1,-1){10}}
\put(50,25){\line(1,1){10}}
\put(36,3){ {\line(0,0){12}}}
\put(30,25){\line(1,-1){10}}
\put(36,15){ {\line(1,1){10}}}
\put(-4,12){$q-y \rightarrow $}
\put(17,0){$q \rightarrow $}
\put(55,23){$\leftarrow q-y-z $}
\end{picture}
\hspace{16.5ex}+
\hspace{10ex}
\begin{picture}(50,30)(0,5)
\put(20,35){\circle{2}}
\put(30,25){\line(1,1){10}}
\put(20,35){\line(1,-1){10}}
\put(40,35){\circle{2}}
\put(50,25){\circle{2}}
\put(26,25){ {\line(1,-1){10}}}
\put(40,15){\line(1,1){10}}
\put(36,3){ {\line(0,0){12}}}
\put(45,12){$\leftarrow q-y$}
\put(45,0){$\leftarrow q$}
\put(-36,23){$q-y-z\rightarrow $}
\end{picture}
\eea
Again, loop graphs appear when products of fields are Wick-ordered. On the ordinary Minkowski spacetime, these
graphs are known as Dyson's double graphs, since they involve both retarded
propagators $\Delta_{ret}$ as well as propagators
$\Delta^{(1)}:=\Delta_++\Delta_-$. On a noncommutative
spacetime they cease  to be equivalent to Feynman graphs. As an example
consider again the fish graph which may in terms of the twisted convolution be
written as
\bea
\lefteqn{\int d^4k\,e^{ikq}\,\int d^4 x\,e^{-ikx}
\,\cdot}\nonumber\\&&\cdot\,
\Big(\,\int dy \,  {\Delta_{ret}}(y)\,
\int dz \, (\,\Delta_-\cdot {\Delta_{ret}}(z)+
 {\Delta_{ret}}\cdot\Delta_+(z)\,)\,
\phi_0(x-y-z)
\\&&+\,
\int dy \,  {\Delta_{ret}}(y)\,
\int dz \, (\,\Delta_-\star  {\Delta_{ret}}(z)+ 
 {\Delta_{ret}}\star  \Delta_+(z)) 
\,\phi_0(x-y-z)\,\,\Big)
\\
&=&
\begin{picture}(60,30)(20,10)
\put(60,35){\circle{2}}
\qbezier(40,15)(20,50)(40,35)
\put(40,35){\line(1,-1){10}}
\put(50,25){\line(1,1){10}}
\put(36,0){ {\line(0,0){15}}}
\put(36,15){ {\line(1,1){10}}}
\put(67,2){$+$}
\end{picture}
\begin{picture}(60,30)(0,10)
\put(0,35){\circle{2}}
\qbezier(20,15)(38,50)(20,35)
\put(16,0){ {\line(0,0){15}}}
\put(0,35){\line(1,-1){10}}
\put(6,25){ {\line(1,-1){10}}}
\put(10,25){\line(1,1){10}}
\end{picture}
+
\begin{picture}(40,30)(0,10)
\put(40,35){\circle{2}}
\qbezier(40,15)(50,60)(60,45)
\qbezier(60,45)(60,32)(50,25)
\put(40,35){\line(1,-1){10}}
\put(36,0){ {\line(0,0){15}}}
\put(36,15){ {\line(1,1){10}}}
\end{picture}
\hspace{7ex}+
\hspace{3ex}
\begin{picture}(60,30)(0,10)
\qbezier(0,45)(10,60)(20,15)
\qbezier(10,25)(0,32)(0,45)
\put(20,35){\circle{2}}
\put(16,0){ {\line(0,0){15}}}
\put(6,25){ {\line(1,-1){10}}}
\put(10,25){\line(1,1){10}}
\end{picture}
\eea
Thus we have again found a planar and a nonplanar contribution. And while in
the planar contribution the time ordering from the retarded propagator may be
absorbed into a Feynman propagator,
\beq
\Delta_{ret}\,(\Delta_++\Delta_-)=-\,i\,\Delta_F^2+i\Delta_-^2
\eeq
this is not the case for the nonplanar contribution, where an additional product
of retarded and advanced propagators appears,
\bea
\Delta_{ret}\star  \Delta_++ \Delta_-\star \Delta_{ret}
&=&-\,i\,\Delta_F\star \Delta_F+\,i\,\Delta_-\star \Delta_-
+\,i\,\Delta_{ret}\star \Delta_{av}
\eea
Precisely this term, which is not present in the context of the modified
Feynman rules, is needed to render the theory unitary~\cite{bdfp}. 
It is absent if spacelike or
lightlike noncommutativity is assumed, since then
$\theta\Delta\star (1-\theta)\Delta=\theta(1-\theta)\cdot \Delta\star\Delta=0$.
We may thus conclude again that only in
these special 
cases, unitarity is compatible with 
the assumption that the time ordering may 
be absorbed in Feynman propagators alone, while the Yang-Feldman approach always renders a unitary theory.


\section{Outlook}

In the above discussion I have not stated what the correct definition of the
Wick product of fields should be. But since some tadpoles remain finite on
noncommutative spacetimes~\cite{seiberg}, we should expect them to be different
from those appearing in the ordinary case. 

As an example let us consider a $3$-fold product of fields. In ordinary quantum
field theory, all tadpoles would be infinite, and the Wick product defined as
\bea
\wick{\phi(x_1)\phi(x_2)\phi(x_3)}
&=&\phi(x_1)\phi(x_2)\phi(x_3)
- \Delta_+(x_1-x_2)\phi(x_3) 
\\\label{ord}
&&
- \Delta_+(x_1-x_3)\phi(x_2)
- \Delta_+(x_2-x_3) \phi(x_1)
\eea
would be well-defined at coinciding points. However, on a noncommutative
spacetime, where we are interested in products such as 
\beq
\phi(q+x_1)\phi(q+x_2)\phi(q+x_3)
\eeq
the subtraction corresponding to the first term in~(\ref{ord}) yields the
following nonplanar expression,
\beq
\int
dk\,\Delta_+(x_1-x_3 - Qk)\,\check\phi(k)
e^{ik(q+x_2)}
\eeq
which remains well-defined at coinciding points $x_i=x$ and therefore does not
need to be subtracted. Moreover, it is nonlocal in the sense that it cannot be
written as a product of a distribution and a field, and therefore, it should
not be subtracted. The precise definition of Wick products where only infinite,
and (in the sense suggested above) local terms are subtracted, as well as the
proof of the adequate Wick theorem are part of our current research. We hope to
be able to shortly produce our results on these questions as well as comment on
the surprising consequences for the so-called ultraviolet/infrared mixing
problem~\cite{seiberg}.


\end{document}